\documentclass[a4paper,11pt]{article}
\usepackage{jheppub}
\usepackage{amsmath}
\usepackage{physics}
\usepackage{slashed}
\usepackage{amssymb}
\usepackage{lineno}
\usepackage{listings}
\usepackage{cleveref}
\crefname{figure}{figure}{figures}
\usepackage{siunitx}
\usepackage{xcolor}
\usepackage{graphicx}
\usepackage{csquotes}
\usepackage{placeins}
\usepackage{subcaption} 

\lstdefinestyle{cppstyle}{
    language=C++,          
    frame=tb,              
    breaklines=true,       
    breakatwhitespace=true, 
    showstringspaces=false, 
    basicstyle=\ttfamily\tiny, 
    keywordstyle=\color{blue}, 
    commentstyle=\color{green!70!black}, 
    stringstyle=\color{red}, 
}
\preprint{MS-TP-25-18}

\title{Conversion of photons to dileptons in the Kroll-Wada and parton shower approaches}

\author[a]{Tom\'a\v{s} Je\v{z}o}
\author[a]{\!\!, Michael Klasen}
\author[b]{and Alexander Puck Neuwirth}

\affiliation[a]{Institut  für  Theoretische  Physik,  Universität  Münster,  Wilhelm-Klemm-Straße 9, 48149 Münster, Germany}
\affiliation[b]{Universit\`a degli Studi di Milano-Bicocca \& INFN Sezione di Milano-Bicocca, Piazza della Scienza 3, Milano 20126, Italy}

\emailAdd{tomas.jezo@uni-muenster.de}
\emailAdd{michael.klasen@uni-muenster.de}
\emailAdd{alexander.neuwirth@unimib.it}

\abstract{
    The study of dileptons in high-energy heavy-ion collisions provides critical insights into the properties of the quark-gluon plasma and the thermal radiation emitted throughout its evolution.
    In the low-mass region, dileptons originate from both direct photon conversion and hadronic decays, with the Kroll-Wada equation traditionally used to relate direct real and direct virtual photon production.
    In this work, we explore the possibility of using parton shower event generators to model this conversion process, leveraging their unitary treatment of internal photon conversions that naturally preserves normalisation, as well as their ability to incorporate higher-order corrections, recoil kinematics, and realistic experimental selection criteria.
    We compare the Kroll-Wada approach to simulations using the {\tt Pythia8} simple shower, the {\tt Vincia} sector shower, and the {\tt POWHEG} shower matched NLO event generator.
    Our results reveal that the parton shower approach offers improved accuracy in describing the dilepton spectrum, particularly towards larger invariant masses where phase-space suppression effects become relevant. %
}

\begin{document}
\maketitle
\flushbottom

\section{Introduction}
\label{sec:intro}

One of the central goals of high-energy heavy-ion physics is to unravel the nature of the quark-gluon plasma (QGP) and its evolution in relativistic collisions.
Among the most valuable probes of this deconfined phase are electromagnetic signals, including real and virtual photons, which escape the medium without undergoing strong final-state interactions\,\cite{Rapp:2009yu,Tserruya:2009zt}.
Dileptons, arising from virtual photon decays, provide a unique window into the space-time evolution of the system, as their invariant mass spectrum directly encodes information about the time evolution of the collision.

The pre-equilibrium effects become evident around $M_{ee} \gtrsim \SI{2}{\giga\electronvolt}$, followed by the expected formation of the QGP in the invariant mass range $\SI{1}{\giga\electronvolt} \lesssim M_{ee} \lesssim \SI{2}{\giga\electronvolt}$, and finally the hadronic phase at later emission stages with $M_{ee} \lesssim \SI{1}{\giga\electronvolt}$ \cite{Rapp:1999ej,Coquet:2021lca}.
However, measuring thermal photon signals in heavy-ion collisions presents considerable challenges, such as the small production cross section due to the electromagnetic coupling, and significant combinatorial and physical backgrounds from hadron decays.
In hadronic collisions, the dielectron mass spectrum in the low and intermediate mass regions is well described by a combination of expected hadronic sources, known as the cocktail\,\cite{PHENIX:2008qav, PHENIX:2009gyd, STAR:2012dzw, ALICE:2018fvj, ALICE:2018gev}.
The low-mass region (LMR, $M_{ee} < \SI{1.1}{\giga\electronvolt} $) is primarily dominated by decays of light-flavoured vector mesons, while the intermediate-mass region (IMR, $\SI{1.1}{\giga\electronvolt} < M_{ee} < \SI{2.7}{\giga\electronvolt}$) is mostly influenced by semileptonic decays of charm and bottom hadrons, which are correlated through flavour conservation.
Despite the significant backgrounds, notable signals of direct real or low-mass virtual photons, over the cocktail, have been observed in heavy-ion collisions at the SPS, RHIC and LHC\,\cite{CERES:1995vll, CERESNA45:1997tgc, CERES:2006wcq, CERESNA45:2002gnc, NA60:2006ymb, NA60:2008ctj, PHENIX:2008uif, PHENIX:2009gyd, STAR:2013pwb, STAR:2015tnn, PHENIX:2014nkk, ALICE:2015xmh}.
These enhancements are well described by exponential distributions in the photons transverse momentum ($p_{T,\gamma}$), suggesting thermal radiation with average (effective) temperatures in the range from $\SIrange{0.2}{0.3}{\giga\electronvolt}$\,\cite{PHENIX:2008uif, PHENIX:2014nkk, ALICE:2015xmh}.
In the intermediate-mass window, an excess over decay dielectrons has only been observed at the SPS\,\cite{NA60:2006ymb, NA60:2008ctj, NA60:2008dcb, NA60:2007lzy}.
To distinguish the subtle thermal photon signal in this range at the LHC, where increased sensitivity to QGP thermal radiation is expected, understanding the dielectron yield originating from the hard production of photons is essential.

A primary method for estimating dileptons in the LMR is the Kroll-Wada formalism \cite{Kroll:1955zu}.
The Kroll-Wada equation was originally developed to describe the photomesonic processes: $\pi^- p \to n e^+ e^-$, $\pi^0 \to \gamma e^+ e^-$ and $\pi^0 \to 2e^+ 2e^-$.
The internal conversion of a photon to a dilepton pair describes the rare decays of vector ($V \to P l^+ l^-$) and pseudoscalar mesons ($P \to V l^+ l^-$) with respect to the radiative decays ($A \to B \gamma$) \cite[Eq.~(3.8)]{Landsberg:1985gaz}.
In the recent years the internal conversion method was applied in a broader context, relating any source of real photons to dileptons \cite{Reygers:2009tm}.
Thereby one obtains a mass distribution of dileptons particularly for the low-mass region, where conventional perturbation theory breaks down due to the extremely low scale.
Often the transverse momentum is significantly larger than the invariant mass $p_T^{ee} \gg M_{ee}$, such that quantum chromodynamics (QCD) resummation allows for the inclusion of a tower of potentially large logarithms \cite{Fai:2003zc,Kang:2008wv,Kang:2009vi}.
The impact of the same $\gamma \to f \bar f$ splittings on dilepton production around the $Z$ mass was studied in \cite{Flower:2022iew}, where it enters as a next-to-next-to-leading order quantum electrodynamics (QED) correction in the soft-photon resummation.

Consequently, measuring dilepton pairs allows to measure the cross section of direct-photon production, which is not overshadowed by the large background of decay photons.
The final dilepton spectrum in the LMR is a combination of direct photons and a cocktail of light and heavy flavour hadron decays \cite{PHENIX:2008qav, ALICE:2018fvj}.
The only fit parameter entering is the ratio of direct to inclusive photons $r$.
It determines the normalization of the direct photon yield, that is given by the Kroll-Wada equation. 

In this work we explore using shower Monte Carlo (MC) event generators as an improvement to the Kroll-Wada equation.
The Kroll-Wada equation is expected to be the lowest order expansion ($\gamma\to e^+ e^-$) of the parton shower (PS) probability.
Such a PS approach offers greater flexibility and precision compared to the Kroll-Wada equation, since it incorporates higher-order contributions in both the hard production process as well as subsequent emissions after the photon splitting, preserves the total rate through a unitary treatment of internal conversions, accounts for sensitivity to the definition of the fiducial region and detector simulations, and allows for direct comparison with experimental data.
In particular we compare the Kroll-Wada equation to the {\tt Pythia8} version 8.313's\footnote{We encountered a bug related to the enhanced radiations that is fixed starting from the indicated version.} simple shower \cite{Bierlich:2022pfr} and {\tt Vincia} dipole shower models \cite{Brooks:2020upa}.
Finally, we demonstrate how the {\tt POWHEG\,BOX\,V2} \cite{Nason:2004rx,Frixione:2007vw} next-to-leading order (NLO) {\tt directphoton} \cite{Jezo:2016ypn} event generator can be used in conjunction with {\tt Pythia} to simulate dileptons without an explicit need to determine the normalization of the theoretical prediction in a fit to data.
This method is universal and can also be applied to other photon production mechanisms, for example direct photon production associated with two jets\footnote{An event generator for this process is also available in {\tt directphotonjj} \cite{Jezo:2024wsc}.}.
The current study intends to complement our recent study in Ref.~\cite{Andronic:2024rfn} in which the same computational approach was used to obtain precise predictions for the $M_{ee}$ spectrum in the IMR.

This paper is structured as follows.
First in \cref{sec:krollwada}, we present the Kroll-Wada equation in its typical form.
We derive it from the Altarelli-Parisi splitting function in \cref{sec:splitting} and in the {\tt Vincia} dipole shower formalism \cite{Brooks:2020upa} in \cref{sec:vincia}.
In \cref{sec:pheno} we compare the PS approach to the Kroll-Wada equation to PHENIX (\cref{sec:phenix}) and ALICE (\cref{sec:alice}) data.
Finally, we conclude in \cref{sec:concl}.

\section{Dielectron pairs from photons}
\label{sec:krollwada}

Photons are an important source of dielectron pairs in LHC experiments. 
One prominent production mechanism is the {\em external photon conversion}, where real photons produced in the collision interact with the detector material and convert into an electron-positron pair via the Bethe-Heitler process \cite{Bethe:1934za}. 
These conversions are well described by QED and are accurately modelled in detector simulations, such as GEANT \cite{GEANT4:2002zbu,Allison:2016lfl}, which include the full material budget and electromagnetic interactions. 
In addition to external conversions, photons also contribute to the dielectron spectrum via {\em internal conversions}, where virtual photons decay into dielectron pairs. 
They emerge in processes that would otherwise produce real photons but instead result in off-shell photons with finite virtuality.
This virtuality directly relates to the invariant mass, $M_{ee}$, of the dielectron pair and approaches zero in the real photon limit. In practice, it never reaches this limit, because the electron mass can no longer be neglected and bounds the invariant mass from below, $M_{ee} \ge 2 m_e$.

Photon-induced dielectron pairs, originating either from external or internal conversion processes, are a key component of the low-mass dielectron spectrum and are essential to model accurately. 
In contrast to external conversions, the modelling of internal conversions does not naturally fit into the now established and indispensable framework of MC event generators. 
In this section, we first review the commonly adopted analytic approach for modelling internal conversions that employs the Kroll-Wada prescription. 
Then we present our new idea on how existing MC--based alternatives achieve the same goal better.

\subsection{Kroll-Wada}
The Kroll-Wada internal electron pair conversion describes a QED process in which an excited system emits an electron-positron pair instead of a real photon.
This process is relevant in the study of meson decays and radiative capture reactions.
The original derivation \cite{Kroll:1955zu} focuses on the theoretical formulation of internal pair production associated with high-energy gamma-ray emission, particularly in pion decays and radiative capture of pions.
It is formulated under the premise that for a given process instead of emitting a real photon a virtual ``nearly real'' photon is emitted.
Then, the virtual photon decays into an electron-positron pair.
This production mode becomes dominant as the energy of the photon approaches zero.
The conversion coefficient $\rho$ is then given as a ratio of the transition probability of emitting a real photon $W_\gamma$ to the transition probability of emitting an electron positron pair $W_\text{pair}$.
The final formula for $\rho$ reads \cite[Eq.~(9)]{Kroll:1955zu}
\begin{equation}
    \begin{aligned}
    \label{eq:kroll_wada_original}
    \rho = \frac{W_\text{pair}}{W_\gamma} &= \frac{2 \alpha}{3 \pi} 
    \int_{2m_e}^E \dd{M_{ee}} \left(\frac{k'}{k}\right) \frac{(E+M_R)^2 + M_R^2 -M_{ee}^2}{(E+M_R)^2 + M_R^2}
    \\&\cdot
    \sqrt{ 1 - \frac{4 m_e^2}{M_{ee}^2}} \left( 1 + \frac{2 m_e^2}{M_{ee}^2}\right)
    \left[ \frac{R_T}{M_{ee}} + \frac{2(E+M_R)^2 M_{ee} }{(2EM_R +E^2 + M_{ee}^2)^2}R_L\right]
    \,.
    \end{aligned}
\end{equation}
Here, $m_e$ is the electron mass, $M_{ee}=m_{\gamma^*}$ is the invariant mass of the dilepton pair, $\alpha$ is the fine structure constant, $E$ is the energy difference between the states before and after the photon emission, $M_R$ is the mass of the recoil system, $k$ and $k'$ are the momenta of the real and virtual photon, respectively.
$R_T$ and $R_L$ are the connected to the transverse and longitudinal polarization vectors of the photon.
Their explicit relation to the spatial components of the (process-dependent) electromagnetic currents $J_\mu(k)$ for $W_\gamma$ and $J_\mu(k')$ for $W_\text{pair}$ is given in \cite[Eq.~(7)]{Kroll:1955zu}.

For practical purposes today, the Kroll-Wada equation is often used to relate the number of real photon to the number of dileptons produced by a virtual photon in the low invariant mass regime.
From \cite[Eq.~(3)]{Reygers:2009tm} we know that the number of $e^+e^-$ pairs per real photon is given in a simplified form by 
\begin{equation}
    \label{eq:kroll_wada_m}
    \frac{1}{N_{\gamma^*}} \frac{\dd N_{ee}}{\dd M_{ee}} = \frac{2 \alpha}{3 \pi} \frac{1}{M_{ee}} \sqrt{ 1 - \frac{4 m_e^2}{M_{ee}^2}} \left( 1 + \frac{2 m_e^2}{M_{ee}^2}\right)S\approx \frac{2 \alpha}{3 \pi}\frac{1}{M_{ee}}
    \,,
\end{equation}
where the only dominant term in $M_{ee}$ is kept. In the equation above, $S=|F(M_{ee}^2)|^2 (1-M_{ee}^2/s)^3$ is a product of a form factor $F$ and a phase space factor. The form factor is for point-like cases exactly equal to one\footnote{In the original Kroll-Wada calculation $F$ is the form factor of a captured or decaying light meson.}. The remaining phase space factor is usually neglected when $M_{ee}^2\ll s = E^2$ or $M_{ee} \ll p_{T,ee}$.
This is similar to the original Kroll-Wada \cref{eq:kroll_wada_original} with a strongly peaked $R_T \sim S$  and thereby negligible $R_L$ for small invariant masses.
In Ref.\,\cite[Eq.~(B1)]{PHENIX:2009gyd} it is 
\begin{align}
    \label{eq:kroll_wada_m2}
    \frac{\dd^2 N_{ee}}{\dd M_{ee}^2} &= \frac{\alpha}{3 \pi} \frac{1}{M_{ee}^2}  \sqrt{ 1 - \frac{4 m_e^2}{M_{ee}^2}} \left( 1 + \frac{2 m_e^2}{M_{ee}^2}\right) \dd N_{\gamma^*}
    \\
    &\approx \frac{\alpha}{3\pi} \frac{1}{M_{ee}^2} \left( 1 - 6 \frac{m_e^4}{M_{ee}^4} - 8 \frac{m_e^6}{M_{ee}^6} \right)  \dd N_{\gamma^*} 
    \label{eq:phenix}
\end{align}
with the photon yields denoted as $\dd N_{\gamma^*}$ and $m_e \ll M_{ee}$.
The factor two difference between \cref{eq:kroll_wada_m} and \cref{eq:kroll_wada_m2} comes from the Jacobian $\dd M_{ee}^2 /\dd M_{ee} = 2M_{ee}$.
The dominant scaling by $1/M_{ee}^2$ could have been guessed just from dimensional considerations and the fact that $M_{ee}$ is the only available significant scale.
The Kroll-Wada equation appears in this simplified form also in \cite[Eq.~(13)]{ALICE:2018fvj}.

Carefully observing \cref{eq:kroll_wada_m} or \eqref{eq:phenix}, we note a fundamental limitation of the Kroll–Wada equation in its lack of {\em unitarity}.
The dilepton pair yield is expressed as the product of the real photon yield and a splitting probability that diverges as the invariant mass $M_{ee}$ approaches zero. 
The inclusion of the electron mass shields the equation from a true divergence, as it enters in the physical lower bound of the integration $\int_{2m_e}^{\sqrt{s}} \mathrm{d}M_{ee}$.
In practice, the integral is regulated by introducing a lower cut off on $M_{ee}$ typically determined by detector resolution, which the predictions are then sensitive to.
For this reason, the normalisation of the $M_{ee}$ spectrum is typically adjusted to experimental data.

In the following subsections we derive the Kroll-Wada equation from a PS perspective and also relate the presence of low invariant mass dileptons to real photon production.
Naturally, an on-shell photon ($M_{\gamma}^2=0$) would not split into a pair of leptons ($M_{ee} \geq 2m_e$) as such splitting would violate energy-momentum conservation.
Nevertheless, a PS will also split an on-shell photon, rebalancing the momentum against a recoiler to enforce its conservation.
In order to understand why this is reasonable consider the following arguments.
One of the roles of PS is to bridge the gap between the high scale of the perturbative hard-scattering process and the low scale non-perturbative phenomena.
Let us denote $Q$ the shower evolution scale.
Whether a photon is on- or off-shell depends on this scale.
This is because the requirement $M_\gamma^2 = 0$ in the hard production process translates to an assumption $M_\gamma^2/Q^2 \approx 0$ when a QED shower evolution is considered.
As $Q$ drops sufficiently low this assumption may no longer be valid,  $M_\gamma^2/Q^2 \neq 0$, and if so the photon being off-shell also needs to be taken into account. 

With the goal of evolving the photon emitted in the hard process from a scale $Q > 2m_e$ down to a lower scale $Q_0$, the PS then proceeds as follows.  
First, it determines whether the photon remains unresolved down to $Q_0 < 2m_e$. 
If no branching occurs above $Q_0$, the photon is treated as on-shell and is not subject to further evolution\footnote{Except for potential momentum reshuffling due to emissions elsewhere in the event.}. 
Otherwise, a branching scale $Q'$ is generated within the range $Q_0 < Q' < Q$, where the photon undergoes a splitting $\gamma \to e^+e^-$. 
The same evolution procedure is then applied recursively to the resulting leptons. 
In this case, the photon initially assumed to be on-shell at scale $Q$ is retroactively assigned a virtuality corresponding to the branching scale $Q'$.

As such parton showers effectively offer an alternative framework for describing internal photon conversions. The derivation of how the PS perspective relates to the Kroll-Wada approach follows.

\subsection{Collinear QED splittings} \label{sec:splitting}
Starting from the familiar Altarelli-Parisi splitting of gluons into quark--anti-quark pairs 
\begin{equation}
    P_{g \to qq}(z) =  T_F ( z^2 + (1-z)^2)
    \,,
\end{equation}
where $z$ is the fraction of gluon's momentum carried by the quark, let us write down the expression for photon splitting into a pair of electrons (cf.~\cite[Eq.~(22)]{deFlorian:2015ujt} referencing \cite{Roth:2004ti})
\begin{equation}
    P_{\gamma \to ee}(z) =  e_e^2 ( z^2 + (1-z)^2)
    \,.
\end{equation}
The splitting is scaled by the electromagnetic charge $e_e^2 = 1$, rather than by the colour structure constant $T_F$.
Interpreting that splitting as a probability
\begin{equation}
\dd \mathcal P_{\gamma \to ee} = \frac{\alpha}{2\pi} \frac{\dd Q^2}{Q^2}  P_{\gamma \to ee}(z) \dd z
\,,
\end{equation}
it is natural to identify the evolution variable with the pair virtuality, $Q^2 = M_{ee}^2$.
This choice is well motivated for splittings of massless gauge bosons into fermion–antifermion pairs since they do not have a soft, but only a collinear singularity \cite{Brodsky:1982gc,Bierlich:2022pfr,Flower:2022iew}.
In this case,
\begin{equation}
\frac{\dd \mathcal P_{\gamma \to ee}}{\dd M_{ee}^2} = \frac{\alpha}{2\pi} \frac{1}{M_{ee}^2}  P_{\gamma \to ee}(z) \dd z 
\,,
\end{equation}
with implicit bounds $2m_e \leq M_{ee} \leq \sqrt{s}$.
We can compute the integral over $z$ explicitly with kinematic bounds 
\begin{align}
\frac{\dd \mathcal P_{\gamma \to ee}}{\dd M_{ee}^2}  
&= \frac{\alpha}{2\pi} \frac{1}{M_{ee}^2}  \int_{y_-}^{y_+}P_{\gamma \to ee}(z) \dd z
\,.
\end{align}
Let's quickly derive these kinematic bounds.
Starting with the momentum $p$ of the photon, the lepton momenta are $p_1$ and $p_2$, with a momentum fraction $z$ and transverse component $k_T$
\begin{align}
    p &= p_1 + p_2
    \,,\\
    p_1 &= z p + k_T
    \,,\\
    p_2 &= (1-z) p - k_T
    \,.
\end{align}
We know that
\begin{align}
    p^2 &= M_{ee}^2 = 2m_e^2 + 2 p_1\cdot{}p_2 =  2m_e^2 + 2  ( z (1-z) p^2 - k_T^2 - (2z-1) p\cdot{}k_T)
    \,,\\
    p_1^2 &= m_e^2 = z^2 p^2 + k_T^2 + 2 z p \cdot{} k_T
    \,,\\
    M_{ee}^2 &=  2m_e^2 + 2  ( z (1-z) M_{ee}^2 - k_T^2 ) - \frac{2z-1}{z}  (m_e^2 - z^2 M_{ee}^2  - k_T^2  )
    \,.
\end{align}
Solving the last expression for $z$ gives
\begin{equation}
 z = \frac 1 2  \pm \frac 1 2  \sqrt{ 1 - 4 \frac{m_e^2}{M_{ee}^2} + 4k_T^2 }
 \,,
\end{equation}
where the bounds correspond to no transverse energy, \textit{i.e.}~$k_T =0$.
In Ref.~\cite[Eq.~(8)]{Kroll:1955zu} the same integration bounds are derived, $y_\pm = \frac{1}{2} \pm \frac{1}{2} \sqrt{1 - \frac{4 m_{e}^{2}}{M_{ee}^{2}}}$.
Thus,
\begin{align}
\frac{\dd \mathcal P_{\gamma \to ee}}{\dd M_{ee}^2}  
&=
\frac{\alpha e_e^2}{3\pi} \frac{1}{M_{ee}^2}  \left(  1  + \frac{m_e^2}{M_{ee}^2}\right)  \sqrt{1 - \frac{4 m_{e}^{2}}{M_{ee}^{2}}}
\\
&\approx \frac{\alpha e_e^2}{3\pi} \frac{1}{M_{ee}^2}  \dd N_{\gamma^*} \left( 1 - \frac{m_e^2}{M_{ee}^2} - 4 \frac{m_e^4}{M_{ee}^4} \right)
\end{align}
is in agreement with \cref{eq:phenix} up to the $m_{e}^2$ terms.
So this simple approximation is already reliably close to Kroll-Wada as long as $M_{ee} \gg m_e$.

It can further be improved by including mass corrections in the splitting function \cite[Eq.~(44)]{Catani:2000ef} 
\begin{equation}
    \label{eq:apm}
    P_{\gamma\to ee}^m = e_e^2 \left( z^2 - (1-z)^2  + 2\frac{m_e^2}{M_{ee}^2}\right)
    \,.
\end{equation}
Here \cref{eq:apm} corresponds to \cite[Eq.~(8)]{Kroll:1955zu} with $z=(1\pm y)/2$.
The $m_e^2$ term is eliminated and
\begin{align}
\label{eq:krollbounds}
\frac{\dd \mathcal P^m_{\gamma \to ee}}{\dd M_{ee}^2}  
&= \frac{\alpha}{2\pi} \frac{1}{M_{ee}^2}  \int_{y_-}^{y_+}P^m_{\gamma \to ee}(z) \dd z
\\
&=\frac{\alpha e_e^2}{3\pi} \frac{1}{M_{ee}^2}  \left(  1  + 2\frac{m_e^2}{M_{ee}^2}\right)  \sqrt{1 - \frac{4 m_{e}^{2}}{M_{ee}^{2}}}
\end{align}
reproduces the Kroll-Wada \cref{eq:kroll_wada_m2} with $S=1$.

\subsection{Parton shower}
\label{sec:vincia}

In the conventional ($1\to2$ splitting) setup, the photon acquires virtuality following the PS algorithm, see e.g.~\cite[Eq.~(101)]{Bierlich:2022pfr},
\begin{equation}
M_{ee}^2 =  \frac{p_T^2}{z(1-z)}
\end{equation}
where $p_T^2$ is the ordering variable of the transverse-momentum ordered shower and $z$ denotes the momentum fraction.
This is a choice made in the simple shower in {\tt Pythia} and has the consequence of the momentum conservation in the collinear limit.
For $\gamma \to f \bar f$ splittings, this is problematic because the production threshold is only recovered in the limit $p_T \to 0$, which is removed by the inevitable infrared cut-off~\cite{Flower:2022iew}.

In more modern dipole showers the PS emissions are instead modelled as $2\to3$ splittings.
One example of such a shower is {\tt Vincia}, which we now turn our attention to.
There the photon splittings are generated with an antenna map in which the evolution variable and the dilepton mass are treated as separate parameters. The virtuality is sampled independently of the $p_T$-like ordering variable, so that the production threshold is respected.
The analytic expressions for the phase space, $\dd \Phi^{\text{FF}}_\text{ant}$, and the antenna function, $\bar{a}_{e/\gamma}^{\text{FF},\gamma}$, for a coherent $IK \to ijk$ splitting, used in {\tt Vincia} can be taken from Ref.~\cite[Eq.~(2.19) and Eq.~(A.22)]{Brooks:2020upa} implementing the sector shower variant:
\begin{align}
    \dd \Phi^{\text{FF}}_\text{ant} = \frac{1}{16 \pi^2} f^{\text{FF}}_\text{Källén} s_{IK} \Theta(\Gamma_{ijk}) \dd y_{ij} \dd y_{jk}  \frac{\dd \phi}{2 \pi}
    \,,\\
    \bar{a}_{e/\gamma}^{\text{FF},\gamma} = \frac{1}{s_{IK}} \frac 1 2 \frac{1}{y_{ij} + 2 \mu_e^2} \left[y_{ik}^2 + y_{jk}^2 + \frac{2 \mu_e^2}{y_{ij} + 2 \mu_e^2}\right]
    \,,
\end{align}
where $s_{IK}=s$, $y_{ij}= s_{ij}/s = 2p_i \cdot p_j/s$ and $\mu_j^2= m_j^2/s$.
Then, with $M_{ee}^2/s = (p_i+p_j)^2/s = y_{ij} + 2 \mu_e^2$ we obtain
\begin{equation}
    \frac{\dd \mathcal P_{\gamma \to ee}^m}{\dd M_{ee}^2} \propto 4 \pi \alpha e_e^2 \frac 1 2 \frac{1}{M_{ee}^2} \left[y_{ik}^2 + y_{jk}^2 + \frac{2 m_e^2}{M_{ee}^2}\right] \frac{1}{16 \pi^2} f^{\text{FF}}_\text{Källén} \Theta(\Gamma_{ijk}) \dd y_{jk}  \frac{\dd \phi}{2 \pi}
\end{equation}
where the $4\pi \alpha e_e^2$ arises from \cite[Eq.~(2.15)]{Brooks:2020upa}.
$f^{\text{FF}}_\text{Källén}  = 1$ as we consider the incoming partons $I$ and $K$ massless.
It is convenient to identify $y_{ik} =z$ and $y_{jk} = 1-z$ in the collinear limit. 
Then the integration bounds become again
\begin{equation}
    0< \Gamma_{ijk} = y_{ij} y_{jk} y_{ik} - y_{jk} \mu_i^2 - y_{ik} \mu_j^2
    \,,
\end{equation}
where we assumed the recoiler to be massless $\mu_k =0$ and solving for $z$ gives
\begin{equation}
    z = \frac 1 2 \pm \frac 1 2 \sqrt{1-\frac{4m_e^2}{M_{ee}^2}}
    \,,
\end{equation}
thereby reproducing the Kroll-Wada \cref{eq:kroll_wada_m2}. 
However, upon examining \cite[Eq.~(2.5)]{Brooks:2020upa} we notice that $y_{jk} + y_{ik} \neq 1$, but rather
\begin{equation}
    1 = y_{ij} + 2 \mu_e^2 + y_{jk} + y_{ik}
    \,.
\end{equation}
Thus, $y_{jk} = 1 - z - M_{ee}^2/s$ modifies also the integration bounds
\begin{align}
   y_\pm =  \frac{\pm\sqrt{\left(M_{ee}^{2} - 2 m_{e}^{2}\right)^{-1} \left(M_{ee}^{2} - s\right) \left(M_{ee}^{4} - 2 M_{ee}^{2} m_{e}^{2} - M_{ee}^{2} s + 6 m_{e}^{2} s\right)} + \left(- M_{ee}^{2} + s\right) }{2 s }
\end{align}
resulting in 
\begin{multline}
\frac{\dd \mathcal P_{\gamma \to ee}}{\dd M_{ee}^2} 
= \sqrt{\frac{M_{ee}^{6} - 2 M_{ee}^{4} m_{e}^{2} - 2 M_{ee}^{4} s + 8 M_{ee}^{2} m_{e}^{2} s + M_{ee}^{2} s^{2} - 6 m_{e}^{2} s^{2}}{M_{ee}^{2} - 2 m_{e}^{2}}}\\
\cdot \frac{\alpha e_e^{2}  \left(- M_{ee}^{8} + 2 M_{ee}^{6} m_{e}^{2} + 2 M_{ee}^{6} s - 5 M_{ee}^{4} m_{e}^{2} s - M_{ee}^{4} s^{2} + 6 m_{e}^{4} s^{2}\right)}{3 \pi M_{ee}^{4} s^{3} \left(- M_{ee}^{2} + 2 m_{e}^{2}\right)}
\,,
\end{multline}
with the simple massless electron limit
\begin{align}
    \label{eq:sscale}
\frac{\dd \mathcal P_{\gamma \to ee}}{\dd M_{ee}^2}  
&=\frac{\alpha e_e^2}{3\pi} \frac{1}{M_{ee}^2}  \left( 1 - \frac{M_{ee}^2}{s}\right)^3 
\,.
\end{align}
This result shows the same $s$ dependent factor as in the original Kroll-Wada publication \cite[Eq.~(13)]{Kroll:1955zu} describing a Dalitz decay with the mass of the pseudo-scalar meson replaced by the partonic centre-of-mass energy.
It also corresponds to \cite[Eq.~(9)]{Kroll:1955zu} with $E^2\to s$, $m\to 0$, $M\to 0$ and $R_T \to  1 - M_{ee}^2/s$.
The differences w.r.t.~\cref{eq:kroll_wada_m2} are part of the often neglected suppressed form factor $S$ in the Kroll-Wada \cref{eq:kroll_wada_m}.

\begin{figure}
    \centering
    \begin{subfigure}{0.45\textwidth}
        \centering
        \includegraphics[height=0.22\textheight]{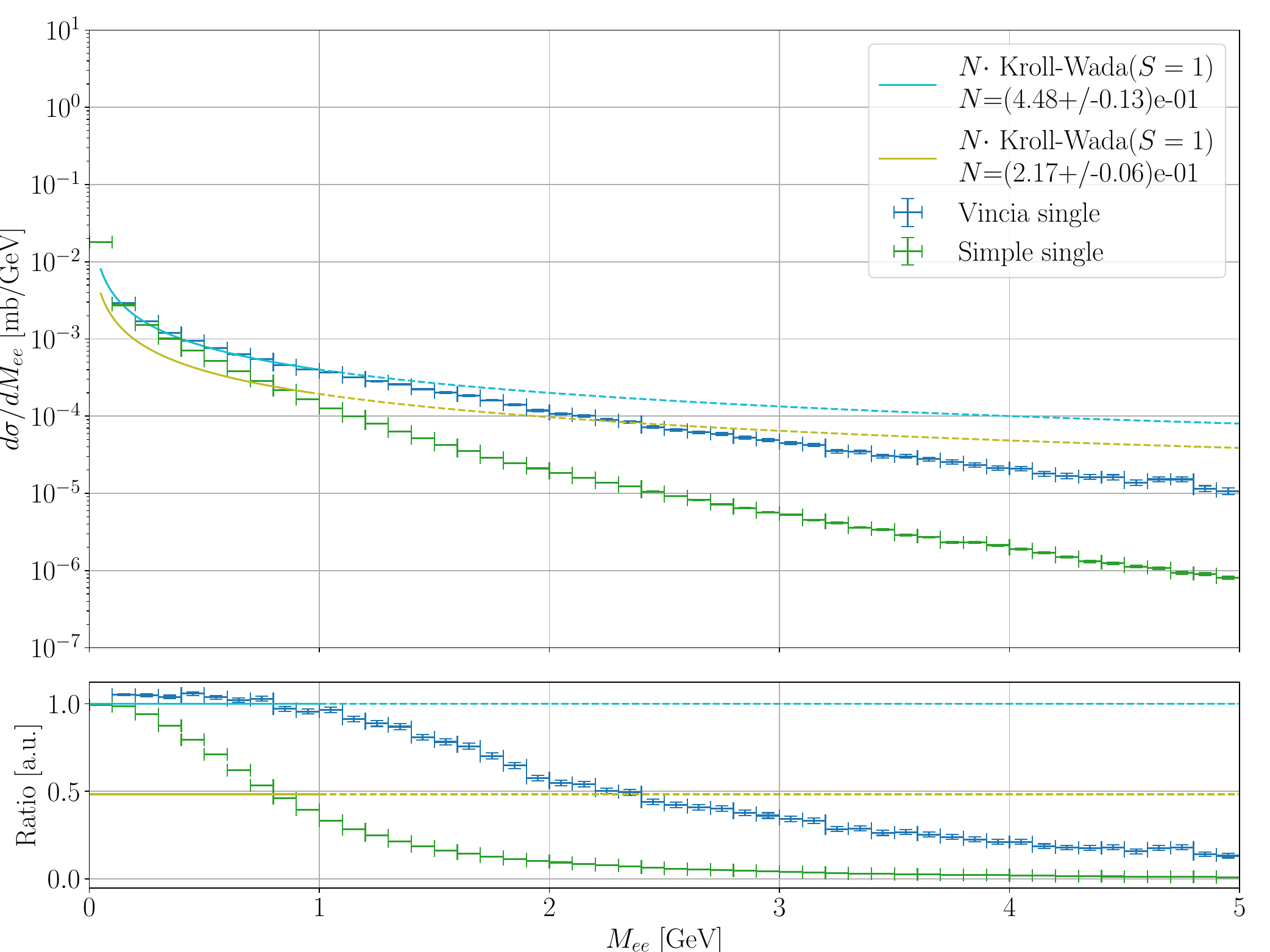}
        \caption{
            The parton shower predictions are normalized to the first bin and the Kroll-Wada equation's normalization is fitted with $S=1$.
        }
        \label{fig:pythia}
    \end{subfigure}
    \hfill
    \begin{subfigure}{0.45\textwidth}
        \centering
        \includegraphics[height=0.22\textheight]{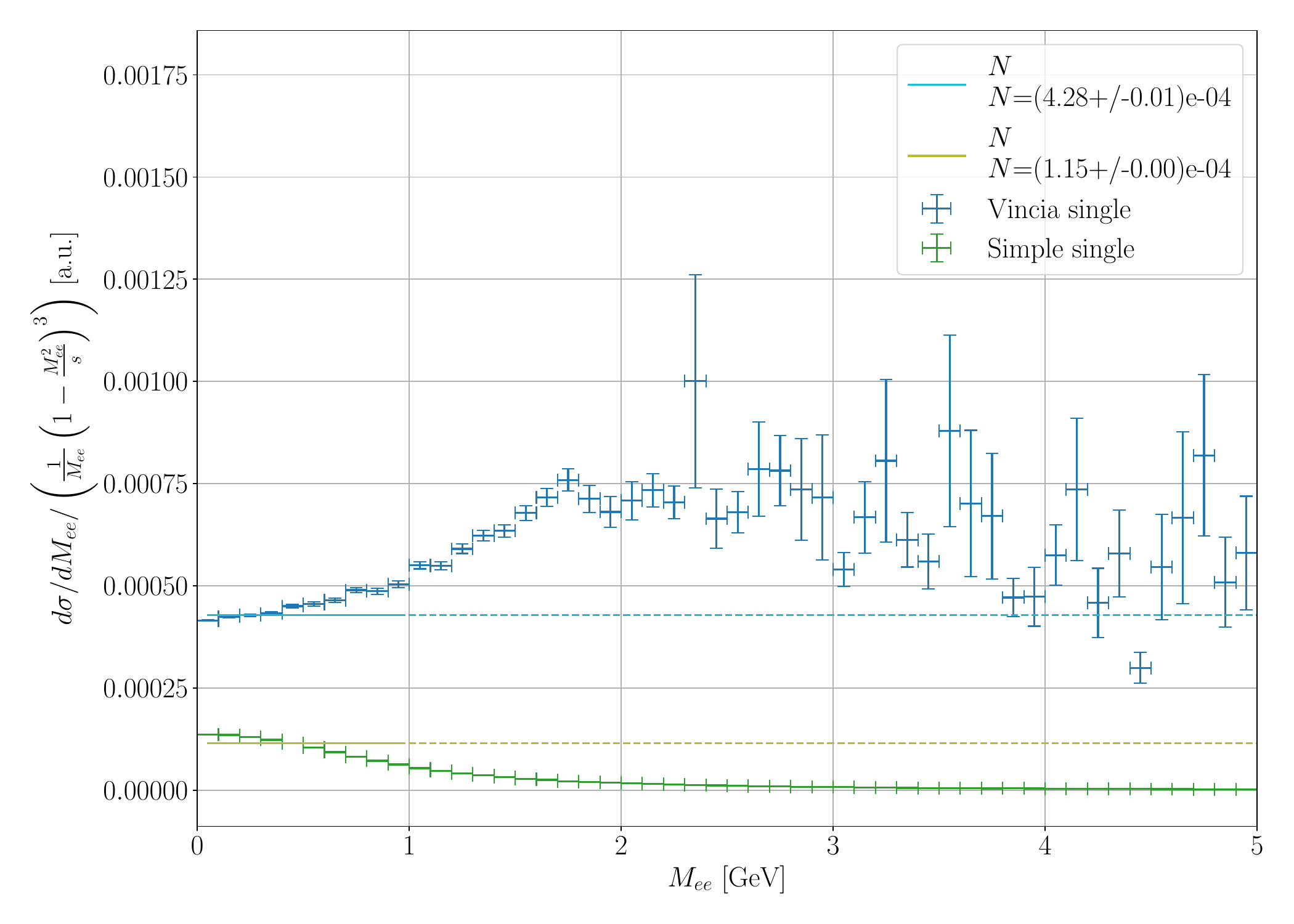}
        \caption{
            The constants are fitted to data where the contribution weights were divided by an event-by-event evaluation of \cref{eq:sscale}.
        }
        \label{fig:pythiakw}
    \end{subfigure}
    \caption{
        Comparison of Kroll-Wada and parton-shower based predictions in pp collisions at $\sqrt{S_{pp}}=\SI{5020}{GeV}$.
        The fits are restricted to $M_{ee} < 1$ GeV (full line) and extrapolated (dashed line).
        All uncertainties are of statistical origin.
        The fit results are obtained using the \texttt{smpl} framework \cite{alexander_puck_neuwirth_2025_15624550}.  
        }
\end{figure}

In \cref{fig:pythia} we examine differences between the Kroll-Wada formula, \cref{eq:kroll_wada_m2}, and various PS predictions.
We consider both the simple shower and the {\tt Vincia} models as implemented in {\tt Pythia} shower MC generator~\cite[Eq.~(101)]{Bierlich:2022pfr}.
Here the {\tt Pythia} simple shower assumes the role of the conventional PS setup, whereas {\tt Vincia} stands in for the dipole-shower approach.
First, an on-shell prompt photon is produced in a hard process of leading order (LO) $pp \to \gamma j$ in {\tt Pythia}.
Then, we shower the event with the two PS limiting both to $\gamma \to ee$ QED splittings in the final state and allowing them to attach only one splitting\footnote{Actually, this restriction is not important, since all further relevant PS steps are prohibited after the first $\gamma \to ee $ splitting. We verified this fact also numerically by allowing shower emissions beyond the first one.}.
So both showered predictions mimic the Kroll-Wada formula in their setup. 
Besides the PS predictions we consider for each a fit to the Kroll-Wada formula in the $M_{ee} < \SI{1}{GeV}$ range, with the normalisation as the only free parameter\footnote{Our loss function is a $\chi^2 = \sum_i (\text{PS}_i-\text{KW}_i)^2/\sigma^2_{\text{PS},i}$, where PS and KW are parton shower and Kroll-Wada predictions, respectively, $\sigma_{\text{PS}}$ is the MC error and $i$ goes over the histograms in the parton shower prediction. Note that the normalisation extracted when minimising this loss function does not necessarily coincide with the physical normalisation. Our goal here is rather to quantify the shape differences.}. 
We find the {\tt Vincia} shower agrees better with Kroll-Wada than the simple shower. Below \SI{1}{GeV} the difference between {\tt Vincia} and Kroll-Wada predictions remains under 10\%, while the simple shower yields a prediction with more pronounced differences in the shape.
Beyond 1 GeV, all three predictions differ significantly from each other. 

In \cref{fig:pythiakw} we show the ratio of our PS predictions to the quantity defined in \cref{eq:sscale}, which is the Kroll-Wada equation in the massless electron limit with $S\neq 1$, but $F=1$, and which we explicitly derived from the {\tt Vincia} framework.
Since both $M_{ee}$ and $s$ in \cref{eq:sscale} depend on the kinematics of the process, \textit{i.e.}~are sampled, a dependence on the partonic centre-of-mass energy $s$ remains, and through the production process in hadronic collisions, so does the dependence on the parton distribution functions.
This sampling of the fluctuating $s$ is presumably responsible for the large statistical uncertainties in the {\tt Vincia} prediction towards larger values of $M_{ee}$, as it is the only difference in the numerator of the ratio in \cref{fig:pythiakw} compared to the absolute prediction in \cref{fig:pythia}. 
By dividing the weight of each event in the analysis with \cref{eq:sscale} one would expect a flat distribution if the events were distributed following this equation.
Therefore, we attempt to fit a constant to the data to check if the equation holds for the PSs.
By comparing the $\chi^2/\text{NDF}$ we find that for small values of $M_{ee}$ {\tt Vincia} matches the above-derived formula, which includes the phase-space factor, much better ($\chi^2/\text{NDF}=\SI{4e2}{}/9$) than the simplified Kroll-Wada equation ($\chi^2/\text{NDF}=\SI{1.9e3}{}/9$).
Nonetheless, in the regime of $M_{ee} < 1$ GeV {\tt Vincia} also agrees well with Kroll-Wada where the $s$ dependence is neglected, \textit{i.e.}~$S=1$.
Our {\tt Vincia} prediction seems to roughly follow \cref{eq:sscale} even for larger values of $M_{ee}$, whereas the simple shower is significantly suppressed. 
It is evident that the extra phase space correction leads to an appreciable departure from the $1/M_{ee}$ since $s \gg M_{ee}^2$ does not hold any more. 

In summary, we discussed how the Kroll-Wada prescription can be derived within the framework of plain collinear or $2 \to 3$ coherent QED splittings available in the {\tt Vincia} MC event generator which implements QED parton showers.
In comparisons to the Kroll-Wada framework we find both {\tt Pythia}'s simple shower and {\tt Vincia} are a decent substitute for modelling ``nearly real'' internal photon conversions. 
For $M_{ee}>\SI{0.5}{GeV}$ {\tt Pythia} prediction develops a modest but clear systematic deviation from Kroll-Wada, whereas {\tt Vincia} remains compatible.
Above \SI{1}{GeV} the inclusion of subleading phase-space effects, often neglected in the Kroll-Wada formalism and available in {\tt Vincia}, become compulsory.
Moreover, on top of the initial $\gamma \to e^+e^-$ splitting, PSs also include further emissions, effectively providing higher order $\alpha$ as well as $\alpha_S$ corrections.
Additionally, they naturally allow for modelling of non-perturbative effects and the inclusion of event selections cuts as well as detector simulations.
In the following sections we confront our predictions to data.

\section{Numeric results} \label{sec:pheno}

In the next sections we compare the Kroll-Wada and the PS predictions to experimental data.
In the first, warm-up, example our goal is to simply demonstrate that LO prompt real photon production scattering process supplemented by a PS can reliably model the direct component of the $M_{ee}$ spectrum in PHENIX data. 
In the second example we step up our game. 
We consider a recent ALICE data and real photon production up to NLO QCD matched to PS, and show that not only we can describe the shape of the data but also its normalisation.
The code used for the PS approach is given in \cref{lst:pythia} and for predictions involving POWHEG in \cref{lst:powheg}.

For a LO EW prediction like the on-shell prompt photon production as in POWHEG the use of the fixed $a(0)\approx 1/137$ would be the common choice.
However, this includes corrections from following photon splittings.
As we add these explicitly by the means of a parton shower the more appropriate EW scheme is $\alpha_{G_\mu} \approx 1/132$ \cite{Denner:2019vbn}.
Both {\tt Pythia} and {\tt Vincia} use one-loop running dynamic coupling $\alpha(Q^2)$ constrained to match $\alpha(M_Z^2)=\alpha_{M_Z} \approx 1/129$.

\FloatBarrier

\subsection{PHENIX} \label{sec:phenix}
The PHENIX experiment has investigated dilepton production in Au+Au and p+p collisions at $\sqrt{S_{NN}} = \SI{200}{GeV}$ \cite{PHENIX:2009gyd}, focusing on the $e^+e^-$ pair continuum across a broad range of masses and transverse momenta.
In the low-mass region, the p+p spectrum is well described by known hadronic decays, whereas Au+Au collisions exhibit a significant enhancement, increasing with centrality.

\begin{figure}
    \begin{center}
    \includegraphics[width=0.8\textwidth]{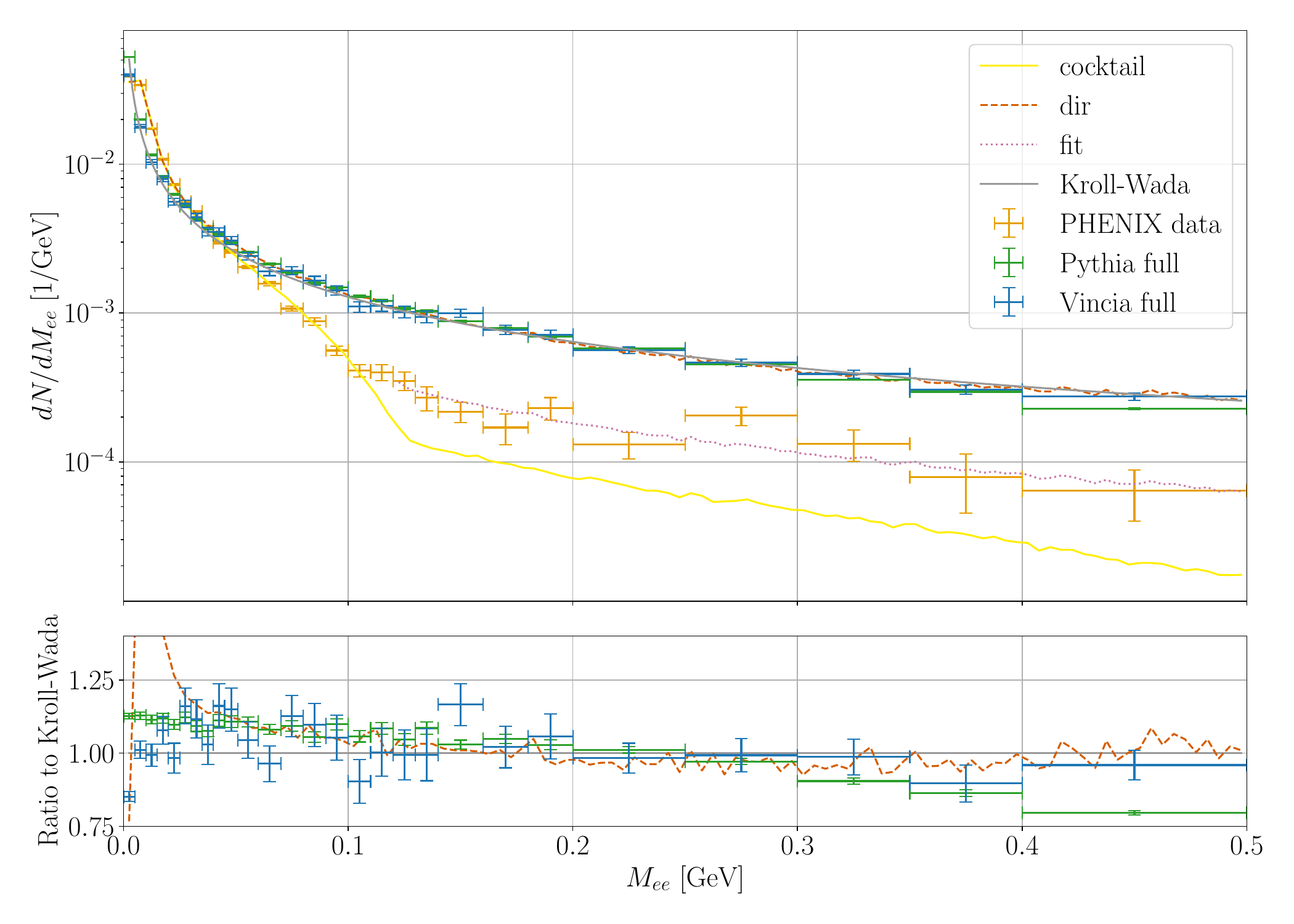}
    \caption{
        All data except the Simple and {\tt Vincia} curves are taken from \cite[Fig. 31]{PHENIX:2009gyd}.
        The uncertainties from the {\tt Pythia} runs are statistical only.
    }
    \label{fig:phenix}
    \end{center}
\end{figure}

In \cref{fig:phenix} we compare the {\tt Pythia} prediction to the Au+Au PHENIX data. 
Firstly, the plot shows the measured data (orange), and the theoretical predictions\footnote{These theoretical predictions were reported in the Au+Au PHENIX publication in Ref.~\cite{{PHENIX:2009gyd}}, where we extracted them from.} decomposed into contributions due to the hadronic cocktail (yellow) and due to the prompt photon production (red, dashed), with arbitrary normalizations. 
The combination of these contributions, with each of the normalization appropriately adjusted to describe the data, is also shown (pink, dotted). 
The scaling factor of the direct component is the ratio of direct to inclusive photons, \textit{i.e.}~$r=N_{\gamma\text{,direct}}/N_{\gamma\text{,inclusive}} = 0.19$ (cf. \cite[Fig. 33]{PHENIX:2009gyd}).
Finally, the plot also displays the prediction from the Kroll-Wada equation (grey) and our prediction obtained with {\tt Pythia}'s simple (green) and {\tt Vincia} (blue) showers. 
Our prediction is based on the LO prompt real photon production process, as implemented in {\tt Pythia}, supplemented by the QED parton shower evolved all the way down to the PS cut-off ({\em full}).
We find the {\tt Vincia} prediction for the invariant mass shape agrees very well with the Kroll-Wada equation, and beyond \SI{0.1}{GeV} also with the {\em direct} component which it is based upon. The prediction based on the simple shower shows a slight shape difference: an increase around \SI{0.1}{GeV} followed by a suppression beyond \SI{0.4}{GeV}. 
We note that this shape difference is consistent with \cref{fig:pythia}, where the $M_{ee}$ range is much larger.
The reason for the direct contribution (red, dashed) not being entirely flat in the ratio panel must be attributed to the applied PHENIX acceptance and detector smearing effects, which are not included in the PS calculations.
Note that the normalization of these predictions is performed analogously to Ref.~\cite{PHENIX:2009gyd} in the $M_{ee}$ range of \SIrange{0.12}{0.3}{GeV}.

We expect both PS calculations to describe the {\em direct} contribution in p+p data, at the same level of accuracy, but are unable to check explicitly as the corresponding decomposition is not available.
As the Kroll-Wada equation does not account for hadronic or nuclear effects, only the overall normalization should differ as compared to Au+Au data, which is nevertheless fitted away. 
An investigation of nuclear effects on the Kroll-Wada equation is beyond the scope of this work.

\FloatBarrier

\subsection{ALICE} \label{sec:alice}
The first measurement of dielectron production at mid-rapidity in pp collisions at \mbox{$\sqrt {S_{pp}} = \SI{7}{TeV}$} with ALICE at the LHC is presented in \cite{ALICE:2018fvj}.
The $e^+e^-$ pairs were studied as a function of invariant mass, transverse momentum, and transverse impact parameter.
The data were compared to a cocktail of expected hadronic sources, with heavy-flavour contributions modelled using {\tt Pythia} and {\tt POWHEG\,BOX}.
In the low-mass region, prompt and non-prompt sources are distinguishable via distance of closest approach of the $e^+e^-$ pair, while in the intermediate-mass region, the total $cc$ and $bb$ cross sections are extracted through a double-differential fit.
The results show good agreement with simulations, though model-dependent differences in heavy-flavour cross sections are observed.
Additionally, the ratio of inclusive to decay photons, measured via virtual direct photons, is consistent with next-to-leading order perturbative QCD predictions.

In this section we compare our predictions for the low-mass region obtained by showering events from prompt (or direct) real photon production. 
We compare them to the Kroll-Wada prediction extracted from Ref.~\cite{ALICE:2018fvj}, which will serve as a proxy of data from the same reference which it describes well.
One important caveat is that $S=1$ has been assumed in the extraction.
Hence comparing against it assumes the phase space factor to be negligible.
This is consistent with the observation in \cref{fig:pythia} for low values of $M_{ee}$.

Besides showering leading order prompt photon events, we will for the first time show the impact of the higher order QCD corrections in the production process.
Higher order corrections in the production process only affect the real photon yield and as such have no impact on the shape of the dilepton spectrum but only on its normalization.
Unlike in the previous figures we do not normalize the parton shower predictions, but compare the absolute cross sections.

\begin{figure}
    \begin{center}
    \includegraphics[width=0.9\textwidth]{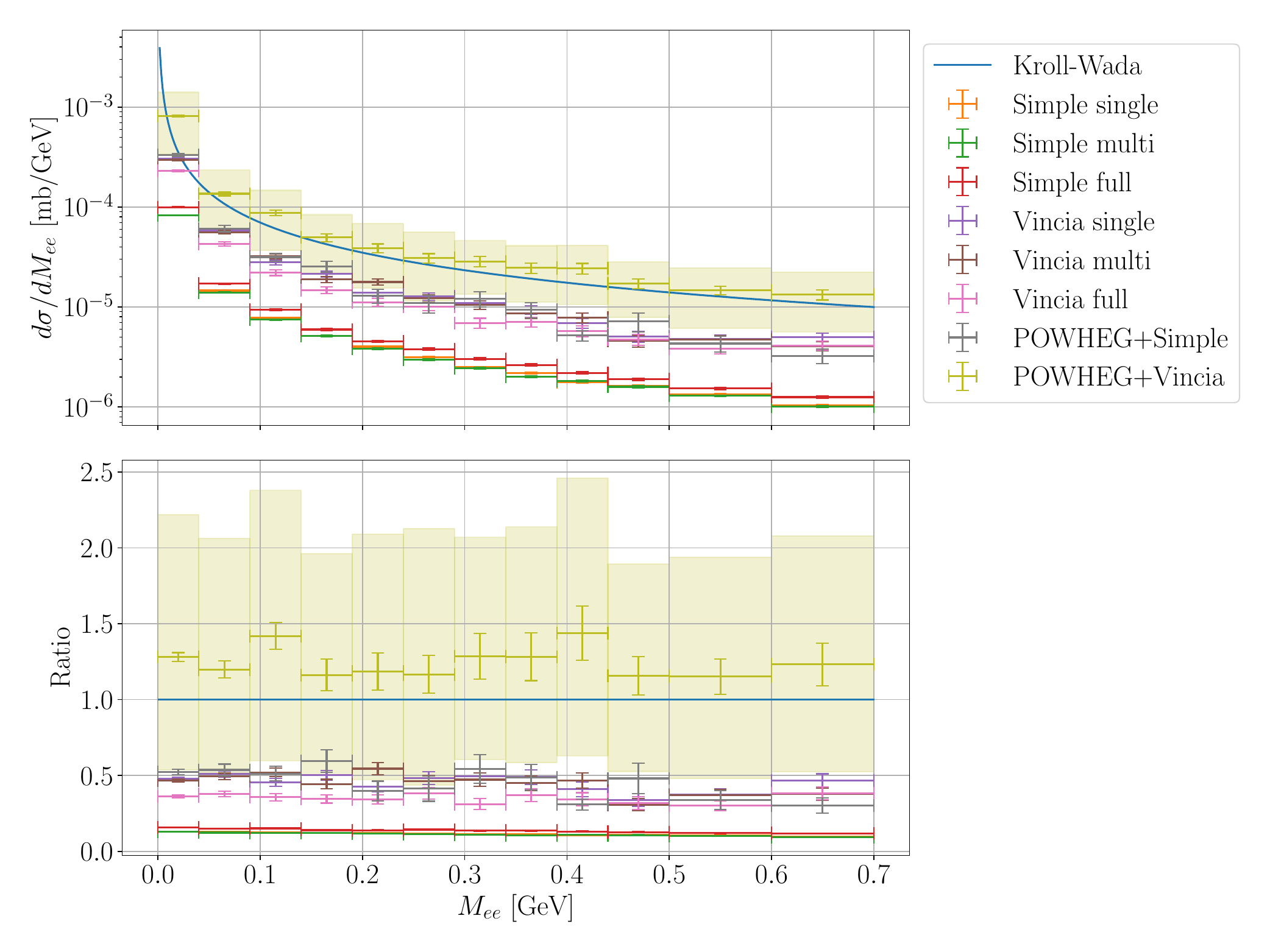}
    \caption{
        The Kroll-Wada function for $pp$ at $\sqrt {S_{pp}} = \SI{7}{TeV}$ used in \cite[Fig. (17)]{ALICE:2018fvj} is shown in comparison to our PS results.
        The analysis cuts applied are $3 < p_{T,ee} < \SI{4}{GeV}$, $|\eta_e| < 0.8$ and $p_{T,e} > \SI{0.2}{GeV}$.
        The {\tt Pythia} and {\tt Vincia} results are given with statistical uncertainties as vertical bars, while the uncertainty band of {\tt POWHEG}+{\tt Vincia} are the typical seven point scale variation.
    }
    \label{fig:alice}
    \end{center}
\end{figure}

\Cref{fig:alice} shows the Kroll-Wada prediction compared to our PS results.
The various {\tt Pythia} predictions, obtained with simple and {\tt Vincia} showers, agree reasonably well with the Kroll-Wada. 
Their normalization can be off by about an order of magnitude, as compared to the Kroll-Wada prediction, in which the normalization has been adjusted to describe the experimental data of \cite{ALICE:2018fvj}.
Their shapes, however, are very similar. 
In the lower ratio panel of \cref{fig:alice}, we see that all parton shower predictions closely follow the Kroll-Wada predictions, showing flat behaviour.
This is consistent with our previous findings.
We also observe a convergence towards the Kroll-Wada result in terms of normalisation, in particular that of {\tt Vincia} is much closer to it than the simple shower. 
We only find small differences in the predictions whether only one ({\em single}) or multiple ({\em multi}) QED shower steps are allowed; and whether or not the QCD shower is switched on ({\em full}).
The difference in normalizations appears to originate from the PS implementations, with {\tt Vincia} yielding consistently larger results by approximately a factor of two, irrespective of whether the photons are generated by {\tt POWHEG} or {\tt Pythia}.
One possible explanation for this difference lies in the direct relationship of the $p_T$ ordering of the simple shower and the photon virtuality $M_{ee}$, which leads to a removal of the important production threshold phase space due to the infrared cut-off.
Another contributing factor could be residual differences in the default parametrizations; although both setups were configured to be consistent to the best of our knowledge, some discrepancies may remain (see \cref{app:shower}).

It must be stressed here that we explicitly prohibit new sources of photons from quarks for the simple shower by setting \verb|TimeShower:QEDshowerByQ| and \verb|SpaceShower:QEDshowerByQ| to \verb|off| and in the {\tt Vincia} shower by pushing the onset of these contributions out of the invariant mass region of interest through \verb|Vincia:QminChgQ = 2.0|.
The parameter \verb|Vincia:QminChgQ| controls the PS's lower cut-off scale of a photon coupling to coloured particles.
Otherwise, with the default of \verb|Vincia:QminChgQ = 0.5|, we find a large increase in the bins beyond $M_{ee} = \SI{0.5}{GeV}$ in the runs where full {\tt Vincia} shower is applied.
This increase is an interplay of the quarks radiating a photon and initial-state radiation, which is not as pronounced in the simple shower\footnote{With the latest version of {\tt Pythia} a combination of its and {\tt POWHEG}'s enhanced radiation features did produce unexpected predictions for the simple shower, which we didn't investigate further. Therefore, {\tt Pythia}'s was disabled for the runs with {\tt POWHEG} events.}.

Finally, when direct photons are produced at NLO QCD using POWHEG's {\tt directphoton} package and subsequently showered with {\tt Vincia}, we find very good agreement in both shape, and within scale uncertainties also in normalization.
At first glance, this might seem unsurprising. 
If the real prompt photon yield is accurate, then the resulting dilepton spectrum should reflect that.
However, this agreement is non-trivial. 
The parton shower is unitary by construction, apart from possible losses due to event migration outside the fiducial region, meaning that the conversion of photons into dileptons preserves normalisation.
That the prediction describes the data without an explicit fit to the overall normalisation both validates the parton shower approach and indicates that the prompt photon yield is significantly more reliable once higher-order QCD corrections are included.\footnote{This feature requires an explicit cut on the transverse momentum of the dilepton pair, since the prompt photon process exhibits a Born-level singularity.}
The remaining large scale uncertainties are a result of probing low values of factorisation and renormalisation scales identified with $p_{T,\gamma}$.

\section{Conclusion} 
\label{sec:concl}
The internal photon conversions into dilepton pairs, a process dominating the $M_{ee}$ spectrum close to zero, are traditionally described using the Kroll-Wada equation.
In this work, we establish that parton showers can provide a reliable alternative for modelling this process, offering a first-principles approach that goes beyond the traditional Kroll-Wada formalism.

We begin by deriving an expression for the probability of internal photon conversion in the presence of a massless recoiler, such as a gluon, as implemented in the {\tt Vincia} parton shower.
This expression reproduces the Kroll-Wada result in the appropriate limit and has the advantage of making the subleading form factor $S$ explicit.
As a result, the parton shower framework predicts a suppression at higher $M_{ee}$, where the Kroll-Wada equation does not fully account for such kinematic configurations.

We then use various PS models to shower direct photon production at LO and NLO and find that {\tt Vincia} reliably matches the shape of the spectrum obtained using the Kroll-Wada equation below $M_{ee} <$ \SI{1}{GeV}. {\tt Pythia}'s simple shower instead starts showing systematics differences above $M_{ee} >$ \SI{0.5}{GeV}.
At higher masses, differences emerge due to the $S$ form factor, which must be taken into account for accurate modelling in this region.
Moreover, thanks to the built-in unitarity of PS, the resulting spectra are correctly normalised provided the input real photon yield is reliable. 
We show that using {\tt POWHEG\,BOX} to determine the real photon yield results in a very good agreement with ALICE data, when showered with {\tt Vincia}, without requiring normalisation corrections.

Furthermore, the parton shower framework naturally allows for the inclusion of higher-order effects and incorporates logarithmic corrections from additional emissions, as well as a more realistic treatment of experimental acceptance. 
Unlike the final-state Kroll–Wada equation, it fully accounts for phase-space cuts, detector effects, and kinematic constraints. 

It would be interesting to explore how the LMR dilepton yield can be combined consistently with the virtual photon yield in the IMR calculated in Ref.~\cite{Andronic:2024rfn}.
Such a procedure could provide further constraints on the normalization of the direct photon contributions to the dilepton mass spectrum.
Also, the comparison of a higher order Kroll-Wada equation to higher order parton showers could provide further validation of the improvements of the PS framework.

\acknowledgments
A.P.N. thanks Christian T.~Preuss for his valuable inputs on the analytic calculation within {\tt Vincia}.
This work has been supported by the BMBF under contract 05P24PMA.
Calculations for this publication were performed on the HPC cluster PALMA II of the University of Münster, subsidised by the DFG (INST 211/667-1).

\paragraph{Open Access.}
This article is distributed under the terms of the Creative Commons Attribution License (CC-BY-4.0), which permits any use, distribution and reproduction in any medium, provided the original authors(s) and source are credited.

\appendix
\section{{\tt Pythia} and {\tt Vincia} code} \label{app:shower}
The code used for the {\tt Pythia} predictions is given below in \cref{lst:pythia}.
The {\tt DIRE} shower was not included in this paper since it is not maintained and would require more adjustments to be used.
\begin{lstlisting}[caption=Pythia code,label=lst:pythia]
// main_kroll_wada.cc

//#define LHE 1
//#define HEPMC 1
#define RIVET 1

//#define SIMPLE  // simple 
//#define VINCIA // VINCIA
//#define DIRE // DIRE

//#define SINGLE // veto everything but first emission
#define ENHANCE // use a2ll to enhance photon -> ee splitting
//#define FULL

//#define PHENIX200
#define ALICE7000
//#define ALICE5020

// Prohibit more than one of the above
#if defined(PHENIX200) + defined(ALICE7000) + defined(ALICE5020) != 1
#error "Exactly one of PHENIX200, ALICE7000, ALICE5020 must be defined"
#endif

#include "Pythia8/Pythia.h"

#ifdef RIVET
#include "Pythia8Plugins/Pythia8Rivet.h"
#include "Rivet/AnalysisHandler.hh"
#endif

#ifdef HEPMC
#include "Pythia8Plugins/HepMC3.h"
#endif

#include <iostream>
#include <chrono>  // For timing

using namespace Pythia8;

class MyUserHooks : public UserHooks {
  // Allow a veto after (by default) first step.
  bool canVetoFSREmission() {
    //exit(-1);
	  return true;
  }
  // Access the event in the interleaved evolution after first step.
  bool  doVetoFSREmission( int sizeOld, const Event& event, int iSys, bool inResonance = false)   {
    if (sizeOld > 7) {
      return true;
    }
    return false;
  }
};

int main(int argc, char* argv[]) {
  int totalIterations = 10000;
  std::string filename = "event";

  if (argc > 1){
	filename = argv[1]; 
  }
  if (argc > 2){
	totalIterations = atoi(argv[2]);
  }

  // Generator.
  Pythia pythia;

  // Set up to do a user veto and send it in.
#ifdef SINGLE
  auto myUserHooks = make_shared<MyUserHooks>();
  pythia.setUserHooksPtr( myUserHooks );
#endif

  //pythia.readString("PDF:pSet = LHAPDF6:NNPDF40_nlo_as_01180/0");
  pythia.readString("Check:event = off");

  // On shell prompt photon
  //pythia.readString("PromptPhoton:all = on");
  // we do not want boxes:
  pythia.readString("PromptPhoton:qg2qgamma = on");
  pythia.readString("PromptPhoton:qqbar2ggamma = on");

  // Kroll-Wada
#ifdef ENHANCE
  pythia.readString("Enhancements:doEnhance = on");
  pythia.readString("EnhancedSplittings:List = { isr:A2AQ=1.,isr:Q2QA=1.,fsr:Q2QA=1.,fsr:A2LL=200.0 }");
#endif
  // disable other lepton sources
  pythia.readString("HadronLevel:all = off");
  pythia.readString("HadronLevel:Decay = off");
  pythia.readString("PartonLevel:MPI = off");
#ifndef FULL
  pythia.readString("PartonLevel:ISR = off");
#endif
  pythia.readString("PartonLevel:FSR = on");
  pythia.readString("PartonLevel:Remnants = off");


#ifdef SIMPLE
  pythia.readString("PartonShowers:model  = 1");
  pythia.readString("SpaceShower:QEDshowerByQ=off");
  pythia.readString("TimeShower:QEDshowerByQ=off");
  // disable QCD shower (=> more QED shower?)
#ifndef FULL
  pythia.readString("TimeShower:QCDshower=off");
  pythia.readString("SpaceShower:QCDshower=off");

  // only photon -> e
  pythia.readString("TimeShower:nGammaToLepton = 1"); 
  pythia.readString("TimeShower:nGammaToQuark = 0");
#endif
#endif
#ifdef DIRE
  pythia.readString("PartonShowers:model=3");
  //pythia.readString("DireSpace:nFinalMax = 0");
  //pythia.readString("DireTimes:nFinalMax = 8");

#ifndef FULL
  pythia.readString("TimeShower:QEDshowerByL=on");
  pythia.readString("TimeShower:QEDshowerByGamma=on");

  pythia.readString("TimeShower:QCDshower=off");
  pythia.readString("SpaceShower:QCDshower=off");
  //pythia.readString("Enhance:Dire_fsr_qed_22->11&11a = 1000");
  //pythia.readString("Enhance:Dire_fsr_qed_22->11&11b = 1000");
#endif
#endif
#ifdef VINCIA
  pythia.readString("PartonShowers:model  = 2");
  pythia.readString("Vincia:mMaxGamma = 10.0"); //1.0 (10.0 default)
  //disable QCD
  pythia.readString("Vincia:QminChgQ = 2.0");
  //pythia.readString("Vincia:convertQuarkToGamma = off");
#ifndef FULL
  pythia.readString("Vincia:nGammaToLepton = 1");
  pythia.readString("Vincia:nGammaToQuark = 0");
  pythia.readString("Vincia:nGluonToQuark = 0");
  pythia.readString("Vincia:convertGluonToQuark = off");
  pythia.readString("Vincia:convertQuarkToGluon = off");
#endif
#endif

  // Use time based seed
  pythia.readString("Random:setSeed = on");
  pythia.readString("Random:seed = 0");

  pythia.readString("PhaseSpace:pTHatMin = 0"); 
  pythia.readString("PhaseSpace:pTHatMax = 10"); 

#ifdef PHENIX200
  pythia.readString("Beams:eCM = 200."); 
#endif
#ifdef ALICE5020
  pythia.readString("Beams:eCM = 5020.");
#endif
#ifdef ALICE7000
  pythia.readString("Beams:eCM = 7000."); 
#endif

#ifdef LHE
  // Create an LHAup object that can access relevant information in pythia.
  LHAupFromPYTHIA8 myLHA(&pythia.process, &pythia.info);

  // Open a file on which LHEF events should be stored, and write header.
  myLHA.openLHEF(filename + ".lhe");

  // Store initialization info in the LHAup object.
  myLHA.setInit();

  // Write out this initialization info on the file.
  myLHA.initLHEF();
#endif

#ifdef RIVET
  //*
  Pythia8Rivet rivet(pythia, filename + ".yoda");
  rivet.ignoreBeams(true);
  rivet.addAnalysis("ALICE_2020_I1797621");
  rivet.addAnalysis("MC_KROLL_WADA");
  rivet.addAnalysis("MC_FSPARTICLES");
  rivet.addAnalysis("MC_DILEPTON");
  rivet.addAnalysis("PHENIX_2019_I1672015");
  rivet.addAnalysis("ATLAS_2014_I1288706");
  rivet.addAnalysis("PHENIX_2009_I838580");
  //*/
#endif


#ifdef HEPMC
  ///*
  // Interface for conversion from Pythia8::Event to HepMC event.
  Pythia8ToHepMC toHepMC(filename + ".hepmc");
  //*/
#endif

   // Extract settings to be used in the main program.
  int    nEvent    = pythia.mode("Main:numberOfEvents");
  int    nAbort    = pythia.mode("Main:timesAllowErrors");

  // Initialization.
  pythia.init();

  // Begin event loop.
  int iAbort = 0;

  // Loop over events.
  for (int i = 0; i < totalIterations; ++i) {

    // Generate event.
    if (!pythia.next()) {

      // If failure because reached end of file then exit event loop.
      if (pythia.info.atEndOfFile()) {
        cout << " Aborted since reached end of Les Houches Event File\n";
        break;
      }

      // First few failures write off as "acceptable" errors, then quit.
      if (++iAbort < nAbort) continue;
      cout << " Event generation aborted prematurely, owing to error!\n";
      break;
    }
#ifdef RIVET
    rivet();
#endif

#ifdef LHE
    // Store event info in the LHAup object.
    myLHA.setEvent();

    // Write out this event info on the file.
    // With optional argument (verbose =) false the file is smaller.
    myLHA.eventLHEF();
#endif
    
#ifdef HEPMC
    toHepMC.writeNextEvent( pythia );
    // */
#endif
  }
#ifdef RIVET
  rivet.done();
#endif

  // Statistics: full printout.
  pythia.stat();

#ifdef LHE
  // Update the cross section info based on Monte Carlo integration during run.
  myLHA.updateSigma();

  // Write endtag. Overwrite initialization info with new cross sections.
  myLHA.closeLHEF(true);
#endif

  // Done.
  return 0;
}


\end{lstlisting}

\section{POWHEG {\tt directphoton} input}
The {\tt POWHEG} input file for the {\tt directphoton} process is given below in \cref{lst:powheg}.
\begin{lstlisting}[language={},caption=POWHEG input card,label=lst:powheg]
numevts 10000 ! number of events to be generated
ih1   1        ! hadron 1 (1 for protons, -1 for antiprotons)
ih2   1        ! hadron 2 (1 for protons, -1 for antiprotons)
ebeam1 3500d0  ! energy of beam 1
ebeam2 3500d0  ! energy of beam 2

! 303400 = NNPDF31_nlo_as_0118
! Remember: you can install pdfs like 'lhapdf install NNPDF31_nlo_as_0118'
! then 'lhapdf update' 
lhans1 27100 ! Pdf set for hadron 1 (LHA numbering)
lhans2 27100 ! pdf set for hadron 2 (LHA numbering)

! for not equal pdf set QCDLambda5
! QCDLambda5 0.22616191135470262

! Parameters to allow or not the use of stored data
use-old-grid    1 ! if 1 use old grid if file pwggrids.dat is present (<> 1 regenerate)
use-old-ubound  1 ! if 1 use norm of upper bounding function stored in pwgubound.dat, if present; <> 1 regenerate

ncall1 100000  ! number of calls for initializing the integration grid
itmx1    5     ! number of iterations for initializing the integration grid
ncall2 100000  ! number of calls for computing the integral and finding upper bound
itmx2    5     ! number of iterations for computing the integral and finding upper bound
foldcsi   2    ! number of folds on csi integration
foldy     5    ! number of folds on  y  integration
foldphi   1    ! number of folds on phi integration
nubound 500000 ! number of bbarra calls to setup norm of upper bounding function
icsimax  1     ! <= 100, number of csi subdivision when computing the upper bounds
iymax    5     ! <= 100, number of y subdivision when computing the upper bounds
xupbound 2d0   ! increase upper bound for radiation generation

bornktmin 2
bornsuppfact 1 ! needed to get photons over a wide range of 100 GeV to over 1 TeV (ATLAS range), not for low energy photons

alphaem_inv  132 ! 1/alphaem

renscfact  1.0d0   ! (default 1d0) ren scale factor: muren  = muref * renscfact 
facscfact  1.0d0   ! (default 1d0) fac scale factor: mufact = muref * facscfact 

#bornonly    1   ! (default 0) if 1 do Born only
#novirtual   1   ! ignore virtuals
#flg_debug   1
emvirtual    1   ! compute soft-virtual QED terms (integrated subtractions)

doublefsr  1  ! redefinition of emitter and emitted, recommended by powheg authors
iupperfsr  1  ! version of upperbounding, 2 ill-defined for massless process

manyseeds  1     ! use pwgseeds.dat for random number input
parallelstage 4
xgriditeration 2

compress_lhe 1
compress_yoda 1

rwl_format_rwgt 1

!! shower/analysis param
NoPythia 0
NoRivet 0
QEDShower 1
QCDShower 1
hadronization 1
mpi 1
pythiaHepMC 1
HepMCEventOutput 0
HepMCEventInput 0
rivetAnalysis "ALICE_2020_I1797621"

rivetAnalysis1 "MC_DILEPTON"
rivetAnalysis2 "PHENIX_2019_I1672015"
rivetAnalysis3 "ATLAS_2014_I1288706"
rivetAnalysis4 "PHENIX_2009_I838580"
rivetAnalysis5 "MC_FSPARTICLES"
rivetAnalysis6 "MC_KROLL_WADA"
rivetWeight 1




! for artificially enhanced statistics. If applied, don't forget to multiply the regular weight with 'sudakovwgt'. 
enhancedradfac 50 ! if > 0 then the splitting kernel used for photon radiation is multiplied by this factor (lhrwgt_id has to be set, too)
#lhrwgt_id 'central'  ! lhrwgt_ reweighting does not work with this version of the code
#lhrwgt_descr 'Central weight, photon splitting enhanced by factor 50' ! lhrwgt_ reweighting does not work with this version of the code
rwl_file '-'
<initrwgt>
<weightgroup name='nominal' combine='None'>
<weight id='default'>default</weight>
<weight id='MUR1.0_MUF1.0_PDF27100'>default</weight>
</weightgroup>
<weightgroup name='scale_variations' combine='None'>
<weight id='MUR2.0_MUF2.0_PDF27100'> renscfact=2d0 facscfact=2d0 </weight>
<weight id='MUR0.5_MUF0.5_PDF27100'> renscfact=0.5d0 facscfact=0.5d0 </weight>
<weight id='MUR1.0_MUF2.0_PDF27100'> renscfact=1d0 facscfact=2d0 </weight>
<weight id='MUR1.0_MUF0.5_PDF27100'> renscfact=1d0 facscfact=0.5d0 </weight>
<weight id='MUR2.0_MUF1.0_PDF27100'> renscfact=2d0 facscfact=1d0 </weight>
<weight id='MUR0.5_MUF1.0_PDF27100'> renscfact=0.5d0 facscfact=1d0 </weight>
</weightgroup>
</initrwgt>
\end{lstlisting}

\bibliography{References}

@software{alexander_puck_neuwirth_2025_15624550,
  author       = {Alexander Puck Neuwirth},
  title        = {APN-Pucky/smpl: v1.5.3},
  month        = jun,
  year         = 2025,
  publisher    = {Zenodo},
  version      = {v1.5.3},
  doi          = {10.5281/zenodo.15624550},
  url          = {https://doi.org/10.5281/zenodo.15624550},
}

@article{Allison:2016lfl,
    author = "Allison, J. and others",
    title = "{Recent developments in Geant4}",
    reportNumber = "FERMILAB-PUB-16-447-CD",
    doi = "10.1016/j.nima.2016.06.125",
    journal = "Nucl. Instrum. Meth. A",
    volume = "835",
    pages = "186--225",
    year = "2016"
}

@article{GEANT4:2002zbu,
    author = "Agostinelli, S. and others",
    collaboration = "GEANT4",
    title = "{GEANT4 - A Simulation Toolkit}",
    reportNumber = "SLAC-PUB-9350, FERMILAB-PUB-03-339, CERN-IT-2002-003",
    doi = "10.1016/S0168-9002(03)01368-8",
    journal = "Nucl. Instrum. Meth. A",
    volume = "506",
    pages = "250--303",
    year = "2003"
}

@inproceedings{Reygers:2009tm,
    author = "Reygers, Klaus",
    collaboration = "PHENIX",
    title = "{Direct Photons at RHIC}",
    booktitle = "{International Conference on the Structure and Interactions of the Photon and 18th International Workshop on Photon-Photon Collisions and International Workshop on High Energy Photon Linear Colliders}",
    eprint = "0908.2382",
    archivePrefix = "arXiv",
    primaryClass = "nucl-ex",
    reportNumber = "HD-KR-002, hd-kr-002",
    pages = "137--143",
    month = "8",
    year = "2009"
}

@article{Kroll:1955zu,
    author = "Kroll, Norman M. and Wada, Walter",
    title = "{Internal pair production associated with the emission of high-energy gamma rays}",
    doi = "10.1103/PhysRev.98.1355",
    journal = "Phys. Rev.",
    volume = "98",
    pages = "1355--1359",
    year = "1955"
}

@article{Brooks:2020upa,
    author = "Brooks, Helen and Preuss, Christian T. and Skands, Peter",
    title = "{Sector Showers for Hadron Collisions}",
    eprint = "2003.00702",
    archivePrefix = "arXiv",
    primaryClass = "hep-ph",
    reportNumber = "CoEPP-MN-20-2, MCNET-20-09",
    doi = "10.1007/JHEP07(2020)032",
    journal = "JHEP",
    volume = "07",
    pages = "032",
    year = "2020"
}

@article{Brodsky:1982gc,
    author = "Brodsky, Stanley J. and Lepage, G. Peter and Mackenzie, Paul B.",
    title = "{On the Elimination of Scale Ambiguities in Perturbative Quantum Chromodynamics}",
    reportNumber = "SLAC-PUB-3011, FERMILAB-PUB-83-040-T",
    doi = "10.1103/PhysRevD.28.228",
    journal = "Phys. Rev. D",
    volume = "28",
    pages = "228",
    year = "1983"
}

@article{Frixione:2007vw,
    author = "Frixione, Stefano and Nason, Paolo and Oleari, Carlo",
    title = "{Matching NLO QCD computations with Parton Shower simulations: the POWHEG method}",
    eprint = "0709.2092",
    archivePrefix = "arXiv",
    primaryClass = "hep-ph",
    reportNumber = "BICOCCA-FT-07-9, GEF-TH-21-2007",
    doi = "10.1088/1126-6708/2007/11/070",
    journal = "JHEP",
    volume = "11",
    pages = "070",
    year = "2007"
}

@article{Jezo:2024wsc,
    author = "Je\v{z}o, Tom\'a\v{s} and Klasen, Michael and Neuwirth, Alexander",
    title = "{Prompt photon production with two jets in POWHEG}",
    eprint = "2409.01424",
    archivePrefix = "arXiv",
    primaryClass = "hep-ph",
    reportNumber = "MS-TP-24-19",
    month = "9",
    year = "2024"
}

@article{Flower:2022iew,
    author = "Flower, Lois and Schoenherr, Marek",
    title = "{Photon splitting corrections to soft-photon resummation}",
    eprint = "2210.07007",
    archivePrefix = "arXiv",
    primaryClass = "hep-ph",
    reportNumber = "IPPP/22/69, MCnet-22-19",
    doi = "10.1007/JHEP03(2023)238",
    journal = "JHEP",
    volume = "03",
    pages = "238",
    year = "2023"
}

@article{PHENIX:2009gyd,
    author = "Adare, A. and others",
    collaboration = "PHENIX",
    title = "{Detailed measurement of the $e^+ e^-$ pair continuum in $p+p$ and Au+Au collisions at $\sqrt{s_{NN}} = 200$ GeV and implications for direct photon production}",
    eprint = "0912.0244",
    archivePrefix = "arXiv",
    primaryClass = "nucl-ex",
    doi = "10.1103/PhysRevC.81.034911",
    journal = "Phys. Rev. C",
    volume = "81",
    pages = "034911",
    year = "2010"
}

@article{Roth:2004ti,
    author = "Roth, Markus and Weinzierl, Stefan",
    title = "{QED corrections to the evolution of parton distributions}",
    eprint = "hep-ph/0403200",
    archivePrefix = "arXiv",
    reportNumber = "MPP-2004-32",
    doi = "10.1016/j.physletb.2004.04.009",
    journal = "Phys. Lett. B",
    volume = "590",
    pages = "190--198",
    year = "2004"
}

@article{ALICE:2018fvj,
    author = "Acharya, Shreyasi and others",
    collaboration = "ALICE",
    title = "{Dielectron production in proton-proton collisions at $ \sqrt{s}=7 $ TeV}",
    eprint = "1805.04391",
    archivePrefix = "arXiv",
    primaryClass = "hep-ex",
    reportNumber = "CERN-EP-2018-102",
    doi = "10.1007/JHEP09(2018)064",
    journal = "JHEP",
    volume = "09",
    pages = "064",
    year = "2018"
}

@article{Kang:2008wv,
    author = "Kang, Zhong-Bo and Qiu, Jian-Wei and Vogelsang, Werner",
    title = "{Low-mass lepton pair production at large transverse momentum}",
    eprint = "0811.3662",
    archivePrefix = "arXiv",
    primaryClass = "hep-ph",
    doi = "10.1103/PhysRevD.79.054007",
    journal = "Phys. Rev. D",
    volume = "79",
    pages = "054007",
    year = "2009"
}

@article{Kang:2009vi,
    author = "Kang, Zhong-Bo and Qiu, Jian-Wei and Vogelsang, Werner",
    editor = "Stankus, Paul and Silvermyr, David and Sorensen, Soren and Greene, Victoria",
    title = "{Low-mass dilepton production in pp and AA collisions}",
    eprint = "0907.4498",
    archivePrefix = "arXiv",
    primaryClass = "hep-ph",
    doi = "10.1016/j.nuclphysa.2009.10.051",
    journal = "Nucl. Phys. A",
    volume = "830",
    pages = "571C--574C",
    year = "2009"
}

@article{deFlorian:2015ujt,
    author = "de Florian, Daniel and Sborlini, Germ\'an F. R. and Rodrigo, Germ\'an",
    title = "{QED corrections to the Altarelli\textendash{}Parisi splitting functions}",
    eprint = "1512.00612",
    archivePrefix = "arXiv",
    primaryClass = "hep-ph",
    reportNumber = "ICAS-03-15, IFIC-15-81",
    doi = "10.1140/epjc/s10052-016-4131-8",
    journal = "Eur. Phys. J. C",
    volume = "76",
    number = "5",
    pages = "282",
    year = "2016"
}

@article{Bierlich:2022pfr,
    author = "Bierlich, Christian and others",
    title = "{A comprehensive guide to the physics and usage of PYTHIA 8.3}",
    eprint = "2203.11601",
    archivePrefix = "arXiv",
    primaryClass = "hep-ph",
    reportNumber = "LU-TP 22-16, MCNET-22-04, FERMILAB-PUB-22-227-SCD",
    doi = "10.21468/SciPostPhysCodeb.8",
    journal = "SciPost Phys. Codeb.",
    volume = "2022",
    pages = "8",
    year = "2022"
}

@article{Landsberg:1985gaz,
    author = "Landsberg, L. G.",
    title = "{Electromagnetic Decays of Light Mesons}",
    doi = "10.1016/0370-1573(85)90129-2",
    journal = "Phys. Rept.",
    volume = "128",
    pages = "301--376",
    year = "1985"
}

@article{Catani:2000ef,
    author = "Catani, Stefano and Dittmaier, Stefan and Trocsanyi, Zoltan",
    title = "{One loop singular behavior of QCD and SUSY QCD amplitudes with massive partons}",
    eprint = "hep-ph/0011222",
    archivePrefix = "arXiv",
    reportNumber = "CERN-TH-2000-336, BI-TP-2000-29",
    doi = "10.1016/S0370-2693(01)00065-X",
    journal = "Phys. Lett. B",
    volume = "500",
    pages = "149--160",
    year = "2001"
}

@article{Andronic:2024rfn,
    author = {Andronic, Anton and Je\v{z}o, Tom\'a\v{s} and Klasen, Michael and Klein-B\"osing, Christian and Neuwirth, Alexander Puck},
    title = "{Di-electron production at the LHC: unravelling virtual-photon and heavy-flavour contributions}",
    eprint = "2401.12875",
    archivePrefix = "arXiv",
    primaryClass = "hep-ph",
    reportNumber = "MS-TP-23-22",
    doi = "10.1007/JHEP05(2024)222",
    journal = "JHEP",
    volume = "05",
    pages = "222",
    year = "2024"
}

@article{Fai:2003zc,
    author = "Fai, George I. and Qiu, Jian-wei and Zhang, Xiao-fei",
    title = "{Full transverse momentum spectra of low mass Drell-Yan pairs at LHC energies}",
    eprint = "hep-ph/0303021",
    archivePrefix = "arXiv",
    doi = "10.1016/j.physletb.2003.05.007",
    journal = "Phys. Lett. B",
    volume = "567",
    pages = "243--250",
    year = "2003"
}

@article{PHENIX:2008uif,
    author = "Adare, A. and others",
    collaboration = "PHENIX",
    title = "{Enhanced production of direct photons in Au+Au collisions at $\sqrt{s_{NN}}=200$ GeV and implications for the initial temperature}",
    eprint = "0804.4168",
    archivePrefix = "arXiv",
    primaryClass = "nucl-ex",
    doi = "10.1103/PhysRevLett.104.132301",
    journal = "Phys. Rev. Lett.",
    volume = "104",
    pages = "132301",
    year = "2010"
}

@article{Coquet:2021lca,
    author = "Coquet, Maurice and Du, Xiaojian and Ollitrault, Jean-Yves and Schlichting, Soeren and Winn, Michael",
    title = "{Intermediate mass dileptons as pre-equilibrium probes in heavy ion collisions}",
    eprint = "2104.07622",
    archivePrefix = "arXiv",
    primaryClass = "nucl-th",
    doi = "10.1016/j.physletb.2021.136626",
    journal = "Phys. Lett. B",
    volume = "821",
    pages = "136626",
    year = "2021"
}

@article{Rapp:1999ej,
    author = "Rapp, R. and Wambach, J.",
    title = "{Chiral symmetry restoration and dileptons in relativistic heavy ion collisions}",
    eprint = "hep-ph/9909229",
    archivePrefix = "arXiv",
    reportNumber = "SUNY-NTG-99-29",
    doi = "10.1007/0-306-47101-9_1",
    journal = "Adv. Nucl. Phys.",
    volume = "25",
    pages = "1",
    year = "2000"
}

@article{Jezo:2016ypn,
    author = {Jezo, Tomas and Klasen, Michael and K\"onig, Florian},
    title = "{Prompt photon production and photon-hadron jet correlations with POWHEG}",
    eprint = "1610.02275",
    archivePrefix = "arXiv",
    primaryClass = "hep-ph",
    reportNumber = "MS-TP-16-23",
    doi = "10.1007/JHEP11(2016)033",
    journal = "JHEP",
    volume = "11",
    pages = "033",
    year = "2016"
}

@article{Rapp:2009yu,
    author = "Rapp, R. and Wambach, J. and van Hees, H.",
    editor = "Stock, R.",
    title = "{The Chiral Restoration Transition of QCD and Low Mass Dileptons}",
    eprint = "0901.3289",
    archivePrefix = "arXiv",
    primaryClass = "hep-ph",
    doi = "10.1007/978-3-642-01539-7_6",
    journal = "Landolt-Bornstein",
    volume = "23",
    pages = "134",
    year = "2010"
}

@article{STAR:2013pwb,
    author = "Adamczyk, L. and others",
    collaboration = "STAR",
    title = "{Dielectron Mass Spectra from Au+Au Collisions at $\sqrt{s_{\rm NN}}$ = 200 GeV}",
    eprint = "1312.7397",
    archivePrefix = "arXiv",
    primaryClass = "hep-ex",
    doi = "10.1103/PhysRevLett.113.022301",
    journal = "Phys. Rev. Lett.",
    volume = "113",
    number = "2",
    pages = "022301",
    year = "2014",
    note = "[Addendum: Phys.Rev.Lett. 113, 049903 (2014)]"
}

@article{STAR:2015tnn,
    author = "Adamczyk, L. and others",
    collaboration = "STAR",
    title = "{Measurements of Dielectron Production in Au$+$Au Collisions at $\sqrt{s_{\rm NN}}$ = 200 GeV from the STAR Experiment}",
    eprint = "1504.01317",
    archivePrefix = "arXiv",
    primaryClass = "hep-ex",
    doi = "10.1103/PhysRevC.92.024912",
    journal = "Phys. Rev. C",
    volume = "92",
    number = "2",
    pages = "024912",
    year = "2015"
}

@article{Tserruya:2009zt,
    author = "Tserruya, Itzhak",
    editor = "Stock, R.",
    title = "{Electromagnetic Probes}",
    eprint = "0903.0415",
    archivePrefix = "arXiv",
    primaryClass = "nucl-ex",
    reportNumber = "WIS-IT-2009-1",
    doi = "10.1007/978-3-642-01539-7_7",
    journal = "Landolt-Bornstein",
    volume = "23",
    pages = "176",
    year = "2010"
}

@article{ALICE:2018gev,
    author = "Acharya, Shreyasi and others",
    collaboration = "ALICE",
    title = "{Dielectron and heavy-quark production in inelastic and high-multiplicity proton\textendash{}proton collisions at $\sqrt {s_{NN}}=$ 13TeV}",
    eprint = "1805.04407",
    archivePrefix = "arXiv",
    primaryClass = "hep-ex",
    reportNumber = "CERN-EP-2018-122",
    doi = "10.1016/j.physletb.2018.11.009",
    journal = "Phys. Lett. B",
    volume = "788",
    pages = "505--518",
    year = "2019"
}

@article{PHENIX:2014nkk,
    author = "Adare, A. and others",
    collaboration = "PHENIX",
    title = "{Centrality dependence of low-momentum direct-photon production in Au$+$Au collisions at $\sqrt{s_{_{NN}}}=200$ GeV}",
    eprint = "1405.3940",
    archivePrefix = "arXiv",
    primaryClass = "nucl-ex",
    doi = "10.1103/PhysRevC.91.064904",
    journal = "Phys. Rev. C",
    volume = "91",
    number = "6",
    pages = "064904",
    year = "2015"
}

@article{PHENIX:2008qav,
    author = "Adare, A. and others",
    collaboration = "PHENIX",
    title = "{Dilepton mass spectra in p+p collisions at s**(1/2) = 200-GeV and the contribution from open charm}",
    eprint = "0802.0050",
    archivePrefix = "arXiv",
    primaryClass = "hep-ex",
    doi = "10.1016/j.physletb.2008.10.064",
    journal = "Phys. Lett. B",
    volume = "670",
    pages = "313--320",
    year = "2009"
}

@article{STAR:2012dzw,
    author = "Adamczyk, L. and others",
    collaboration = "STAR",
    title = "{Di-electron spectrum at mid-rapidity in $p+p$ collisions at $\sqrt{s} = 200$ GeV}",
    eprint = "1204.1890",
    archivePrefix = "arXiv",
    primaryClass = "nucl-ex",
    doi = "10.1103/PhysRevC.86.024906",
    journal = "Phys. Rev. C",
    volume = "86",
    pages = "024906",
    year = "2012"
}

@article{CERES:1995vll,
    author = "Agakichiev, G. and others",
    collaboration = "CERES",
    title = "{Enhanced production of low mass electron pairs in 200-GeV/u S - Au collisions at the CERN SPS}",
    reportNumber = "CERN-PPE-95-026, CERN-PPE-95-26",
    doi = "10.1103/PhysRevLett.75.1272",
    journal = "Phys. Rev. Lett.",
    volume = "75",
    pages = "1272--1275",
    year = "1995"
}

@article{CERESNA45:1997tgc,
    author = "Agakichiev, G. and others",
    collaboration = "CERES/NA45",
    title = "{Low mass e+ e- pair production in 158/A-GeV Pb - Au collisions at the CERN SPS, its dependence on multiplicity and transverse momentum}",
    eprint = "nucl-ex/9712008",
    archivePrefix = "arXiv",
    doi = "10.1016/S0370-2693(98)00083-5",
    journal = "Phys. Lett. B",
    volume = "422",
    pages = "405--412",
    year = "1998"
}

@article{CERES:2006wcq,
    author = "Adamova, D. and others",
    collaboration = "CERES",
    title = "{Modification of the rho-meson detected by low-mass electron-positron pairs in central Pb-Au collisions at 158-A-GeV/c}",
    eprint = "nucl-ex/0611022",
    archivePrefix = "arXiv",
    doi = "10.1016/j.physletb.2008.07.104",
    journal = "Phys. Lett. B",
    volume = "666",
    pages = "425--429",
    year = "2008"
}

@article{CERESNA45:2002gnc,
    author = "Adamova, D. and others",
    collaboration = "CERES/NA45",
    title = "{Enhanced production of low mass electron pairs in 40-AGeV Pb - Au collisions at the CERN SPS}",
    eprint = "nucl-ex/0209024",
    archivePrefix = "arXiv",
    doi = "10.1103/PhysRevLett.91.042301",
    journal = "Phys. Rev. Lett.",
    volume = "91",
    pages = "042301",
    year = "2003"
}

@article{NA60:2006ymb,
    author = "Arnaldi, R. and others",
    collaboration = "NA60",
    title = "{First measurement of the rho spectral function in high-energy nuclear collisions}",
    eprint = "nucl-ex/0605007",
    archivePrefix = "arXiv",
    doi = "10.1103/PhysRevLett.96.162302",
    journal = "Phys. Rev. Lett.",
    volume = "96",
    pages = "162302",
    year = "2006"
}

@article{NA60:2008ctj,
    author = "Arnaldi, R. and others",
    collaboration = "NA60",
    title = "{NA60 results on thermal dimuons}",
    eprint = "0812.3053",
    archivePrefix = "arXiv",
    primaryClass = "nucl-ex",
    doi = "10.1140/epjc/s10052-009-0878-5",
    journal = "Eur. Phys. J. C",
    volume = "61",
    pages = "711--720",
    year = "2009"
}

@article{ALICE:2015xmh,
    author = "Adam, Jaroslav and others",
    collaboration = "ALICE",
    title = "{Direct photon production in Pb-Pb collisions at $\sqrt{s_{NN}} =$ 2.76 TeV}",
    eprint = "1509.07324",
    archivePrefix = "arXiv",
    primaryClass = "nucl-ex",
    reportNumber = "ALICE-PUBLIC-2015-007, CERN-PH-EP-2015-254",
    doi = "10.1016/j.physletb.2016.01.020",
    journal = "Phys. Lett. B",
    volume = "754",
    pages = "235--248",
    year = "2016"
}

@article{NA60:2008dcb,
    author = "Arnaldi, R and others",
    collaboration = "NA60",
    title = "{Evidence for the production of thermal-like muon pairs with masses above 1-GeV/c**2 in 158-A-GeV Indium-Indium Collisions}",
    eprint = "0810.3204",
    archivePrefix = "arXiv",
    primaryClass = "nucl-ex",
    doi = "10.1140/epjc/s10052-008-0857-2",
    journal = "Eur. Phys. J. C",
    volume = "59",
    pages = "607--623",
    year = "2009"
}

@article{NA60:2007lzy,
    author = "Arnaldi, R. and others",
    collaboration = "NA60",
    title = "{Evidence for radial flow of thermal dileptons in high-energy nuclear collisions}",
    eprint = "0711.1816",
    archivePrefix = "arXiv",
    primaryClass = "nucl-ex",
    doi = "10.1103/PhysRevLett.100.022302",
    journal = "Phys. Rev. Lett.",
    volume = "100",
    pages = "022302",
    year = "2008"
}

@article{Nason:2004rx,
    author = "Nason, Paolo",
    title = "{A New method for combining NLO QCD with shower Monte Carlo algorithms}",
    eprint = "hep-ph/0409146",
    archivePrefix = "arXiv",
    reportNumber = "BICOCCA-FT-04-11",
    doi = "10.1088/1126-6708/2004/11/040",
    journal = "JHEP",
    volume = "11",
    pages = "040",
    year = "2004"
}

@article{Bethe:1934za,
    author = "Bethe, H. and Heitler, W.",
    title = "{On the Stopping of fast particles and on the creation of positive electrons}",
    doi = "10.1098/rspa.1934.0140",
    journal = "Proc. Roy. Soc. Lond. A",
    volume = "146",
    pages = "83--112",
    year = "1934"
}

@article{Denner:2019vbn,
    author = "Denner, Ansgar and Dittmaier, Stefan",
    title = "{Electroweak Radiative Corrections for Collider Physics}",
    eprint = "1912.06823",
    archivePrefix = "arXiv",
    primaryClass = "hep-ph",
    reportNumber = "FR-PHENO-019",
    doi = "10.1016/j.physrep.2020.04.001",
    journal = "Phys. Rept.",
    volume = "864",
    pages = "1--163",
    year = "2020"
}

\end{document}